\documentclass[useAMS, usenatbib]{mn2e}

\usepackage{graphicx}

\title[Intrinsic velocity field in the M87 jet]{Determination of the intrinsic velocity field in the M87 jet}
\author[C.-C. Wang and H.-Y. Zhou]{Chun-Cheng~Wang$^{1,2}$\thanks{E-mail:
ccwang@ustc.edu.cn} and Hong-Yan~Zhou$^{1,2}$\\
$^{1}$Center for Astrophysics, University of Science and
Technology of China,
 Hefei, Anhui 230026, P.R.~China\\
$^{2}$Joint Institute for Galaxy and Cosmology (JOINGC), SHAO and USTC, CAS}
\begin{document}

\date{Accepted 2008 December 30. Received 2008 December 2; in original form 2007 December 4}

\pagerange{\pageref{firstpage}--\pageref{lastpage}} \pubyear{2009}

\maketitle

\label{firstpage}

\begin{abstract}
A new method to estimate the Doppler beaming factor of relativistic
large-scale jet regions is presented. It is based on multiwaveband
fitting to radio-to-X-ray continua with synchrotron spectrum models.
Combining our method with available observational data of proper
motions, we derive the intrinsic velocity as well as the viewing
angles to the line of sight for eight knotty regions down the M87
jet. The results favor the 'modest beaming' scenario along the jet,
with Doppler factors varying between $\sim2-5$. The inner jet of M87
suffers sharp deceleration, and the intrinsic speed remains roughly
constant down the outer jet. The orientation of the inner jet
regions is fully consistent with the result of 10\degr-19\degr to
the line of sight suggested by previous {\sl Hubble Space Telescope}
({\sl HST}) proper motion studies of the M87 jet. The outer jet,
however, shows systematic deflection off the inner jet to much
smaller inclination ($\theta\ll10$\degr). Further calculation of
knot~A suggests this deflection can be regarded as evidence that the
outer jet suffers some departure from equipartition. The nucleus
region of the M87 jet should have a viewing angle close to its first
knot HST-1, i.e. $\theta\sim15$\degr, which favors the idea that M87
may be a misaligned blazar. This work provides some hints about the
overall dynamics of this famous extragalactic jet.
\end{abstract}


\section{Introduction}

Since the discovery of the first extragalactic large-scale jet in
a nearby active galaxy M~87 \citep{Curtis}, intense efforts are
made to explore the underlying physics of the relativistic jets
during the past decades. The related topics concern about the
matter content, particle acceleration, collimation mechanism and
magnetic field configuration inside the jet flow. Among these
problems, one of the fundamental issues is the intrinsic velocity
field of the jet. Lack of an effective method to determine the
speed and orientation of the successive regions in large-scale
jet, it is difficult to constrain further the mechanism on overall
dynamics of the bulk jet flow.

In this work, a new method is presented to estimate the Doppler
beaming factor in a region of large-scale jet. This method is
based on the spectral fit of the multi-waveband continua with
standard synchrotron radiation models. Combining the data of
proper motions, we readily derive the intrinsic bulk velocity as
well as the angle to line-of-sight of the jet flow. Making
application of this new method, we successfully obtain the
intrinsic velocity field in M~87 jet. Our result agrees well with
that of the new variability monitoring program toward the same
object \citep{Harris06}, which indicates the reliability of our
approach. The method is simple and only requires non-simultaneous
spectral energy distribution (SED) data sets, therefore is hopeful
to be of general interest for those synchrotron-dominated jets.

For the rest of this paper, we firstly make synchrotron spectrum
model fitting to radio-to-X-ray SED of M87 jet in \S2, then
introduce the new method and derive the distribution of Doppler
beaming factor along M~87 jet in \S3. Coupling the data of proper
motions, we obtain the intrinsic velocity field of M87 jet in \S4.
Related discussions are also presented in this section. Finally,
the main results are summarized in \S5.

\section[]{Synchrotron Spectrum Model Fitting to M87 Jet}

\subsection[]{The Synchrotron Radiation Models}

The high spatial resolution of {\it Chandra} X-ray observatory
(FWHM = 0.5\arcsec) has opened a new era on the research of
extragalactic large-scale jets at high energy waveband. At the
time of preparing this paper, 91 radio-loud AGNs are reported
detection of X-ray counterparts of radio jets on large-scales (see
also {\tt http://hea-www.harvard.edu/XJET/}). The number of
detected X-ray jets increases rapidly, suggesting that X-ray
emission from jets is a common feature in radio galaxies and
quasars. If combined with observational data in radio and optical
band (sometimes also with available infrared and ultraviolet
data), we can construct broadband SED and carry out multi-waveband
modelling of the jet regions. Such a procedure may provide us with
useful information of jet physics.

Polarization observations already confirm the synchrotron nature of
the radio-optical emissions from extragalactic jets (e.g.,
\citealt{Perlman99} for M~87 jet). The X-ray emission process of
large-scale jets, however, remains as an open question. For nearby
FR~I radio galaxies, the radio-to-X-ray SED from the large-scale
knots is consistent with a single smoothly broken power-law
spectrum, and the expected inverse-Compton scattering (IC) fluxes of
seed photons (jet synchrotron photons or cosmic microwave background
photons) always underpredict the observed X-ray fluxes. This implies
the synchrotron origin of the X-ray emissions observed in FR~I jets
(e.g., \citealt{Marshall02} and \citealt{Wilson02} for M~87,
\citealt*{Hardcastle01} for 3C~66B). The recent statistical research
of a large X-ray jet sample \citep{Kataoka05} also favors this
physical picture. On the other hand, the X-ray emissions from
large-scale quasar and FR~II jets, however, are usually interpreted
as inverse-Compton scattering of CMB photons within a highly
relativistic jet \citep{Harris02}.

As discussed by \citet{Schwartz02}, the IC/CMB model for the X-ray
emissions of large-scale quasar jets predicts two main differences
between nearby and distant objects. Due to the enhancement of the
CMB energy density with redshift by a factor of $(1+z)^4$, the
ratio of X-ray to radio fluxes of the resolved parts of the quasar
jets should increase in distant sources compared to nearby ones.
The other effect is that we should observe some systematic
flattening of the optical-to-X-ray spectral index ($\alpha_{\rm
OX}$) in distant unresolved quasar cores compared to low-redshift
quasars, because of the increasing contribution from the
unresolved portions of the jet to the X-ray flux of the quasar
core.

We notice, however, the recent observational efforts of
\citet{Lopez06} to search the latter effect. They carry out a {\it
Chandra} snapshot survey of representative high-redshift radio-loud
quasars ($z\approx3.5-4.7$) selected from the Parkes-MIT-NRAO
sample. The survey does not detect any systematic flattening of the
optical-to-X-ray spectral index ($\alpha_{\rm OX}$) of the
unresolved quasar cores compared to low-redshift quasars. The
results of \citet{Lopez06} suggest that kiloparsec-scale X-ray
emission of quasar jet is not dominated by inverse Compton
scattering of CMB seed photons off jet electrons. Meanwhile, the
IC/CMB model also meets difficulty in explaining the X-ray emissions
from several FR~II jets (e.g., \citealt{Kraft05} for 3C~403).  It is
also worthy to notice the most recent discovery of the X-ray
counterjet in the FR~II source 3C~353 detected by {\sl Chandra}
\citep{Kataoka08}. The research of \citet{Kataoka08} suggests that
this detection is inconsistent with the IC/CMB model, and instead
implies a synchrotron origin of the X-ray jet photons. Thus, even
for powerful quasar/FR~II jets, the synchrotron radiation mechanism
for X-ray photons keeps as a plausible interpretation.

In general, synchrotron origin for X-ray emission of large-scale jet
faces much more strict challenge in particle acceleration than
IC/CMB model does. Firstly, the particle acceleration in synchrotron
X-ray jets has to be fast enough to generate ultra-relativistic
electrons with maximum energies $\sim10-100$~TeV (i.e., one to two
orders of magnitude higher than that required in the IC/CMB model).
Secondly, the electron acceleration processes in synchrotron X-ray
jets have to operate continuously within the whole volumes of the
jets. Such a strict requirement that synchrotron X-ray jets should
work as powerful particle accelerators is naturally supported by the
idea that large-scale jets are very hopeful sites for yielding Ultra
High Energy Cosmic Ray particles (UHECRs, see e.g. \citealt{Casse05}
and references therein).

Based on the observations mentioned above, we choose synchrotron
radiation model for our multi-waveband research. Currently, there
are three standard synchrotron spectrum models available for
statistical fit of SED: (1) The \citet{Jaffe73} model (hereafter
JP), which assumes the pitch angle distribution of relativistic
electrons is continuously isotropized after an initial injection
of single power-law electron energy distribution. (2) The
Kardashev-Pacholczyk model (\citealt{Kardashev62} and
\citealt{Pacholczyk70}, hereafter KP), which allows only the
evolution of high energy tail of electron energy distribution
following an initial single-power law electron injection. Since no
pitch-angle scattering of the radiating electrons is allowed, it
is unlikely to be the real case. This model, however, can give
good fits to the observational data in many cases, therefore is
relatively common in the jet research community. (3) The
Continuous Injection model (\citealt{Heavens87} and
\citealt{Meisenheimer89}, hereafter CI), which allows continuous
electron injection with single power-law energy distribution into
the emission region.

The spatial extent of the X-ray emitting regions in some FR~I jets
resolved by {\sl Chandra} implies that continuous acceleration and
injection of high-energy electrons may occur in these objects. On
the other hand, recent work of \citet{Perlman05} on M87 jet suggests
that JP model underpredicts the X-ray flux by many orders of
magnitude and the theoretical slope at X-ray energies is much larger
than those observed. Therefore we do not consider the JP model in
our work. All of the multi-band fits are made using KP and CI
models. For all of the synchrotron spectrum models above, the
dominated cooling process of high-energy electrons responsible for
X-ray emissions should be synchrotron radiative energy loss.

\subsection[]{Result of Multi-waveband Fit}
M~87 is one of the nearby FR~I radio galaxies (distance = 16~Mpc,
with a scale of 1\arcsec = 77.6~pc). The rich observational data
of its famous jet enable us to do further multi-band fit. The
program of synchrotron spectrum fit we use is written by
C.~Carilli and J.P.~Leahy (\citealt{Carilli91};
\citealt{Leahy91}). The fit is done in two steps. In step~1, we
get an initial guess of the position of the break frequency by
employing a Myers and Spangler test \citep{Myers85} on the
multi-waveband data sets. Then in step~2, a Marquardt non-linear
$\chi^2$ test is done numerically \citep{Press87} to reach the
best fitting of the SED data. The tests are model dependent. The
electrons responsible for higher frequency synchrotron radiation
suffer more rapid synchrotron loss, thus there is a spectra
steepening toward the high frequency band. For CI model, there is
a break in spectral index of $\Delta\alpha =0.5$ between the low
and high frequency spectra. In the case of KP model,
$\Delta\alpha$ depends on the index of electron energy
distribution.

\begin{figure*}
\includegraphics[angle=-90, totalheight=100mm, keepaspectratio=true]{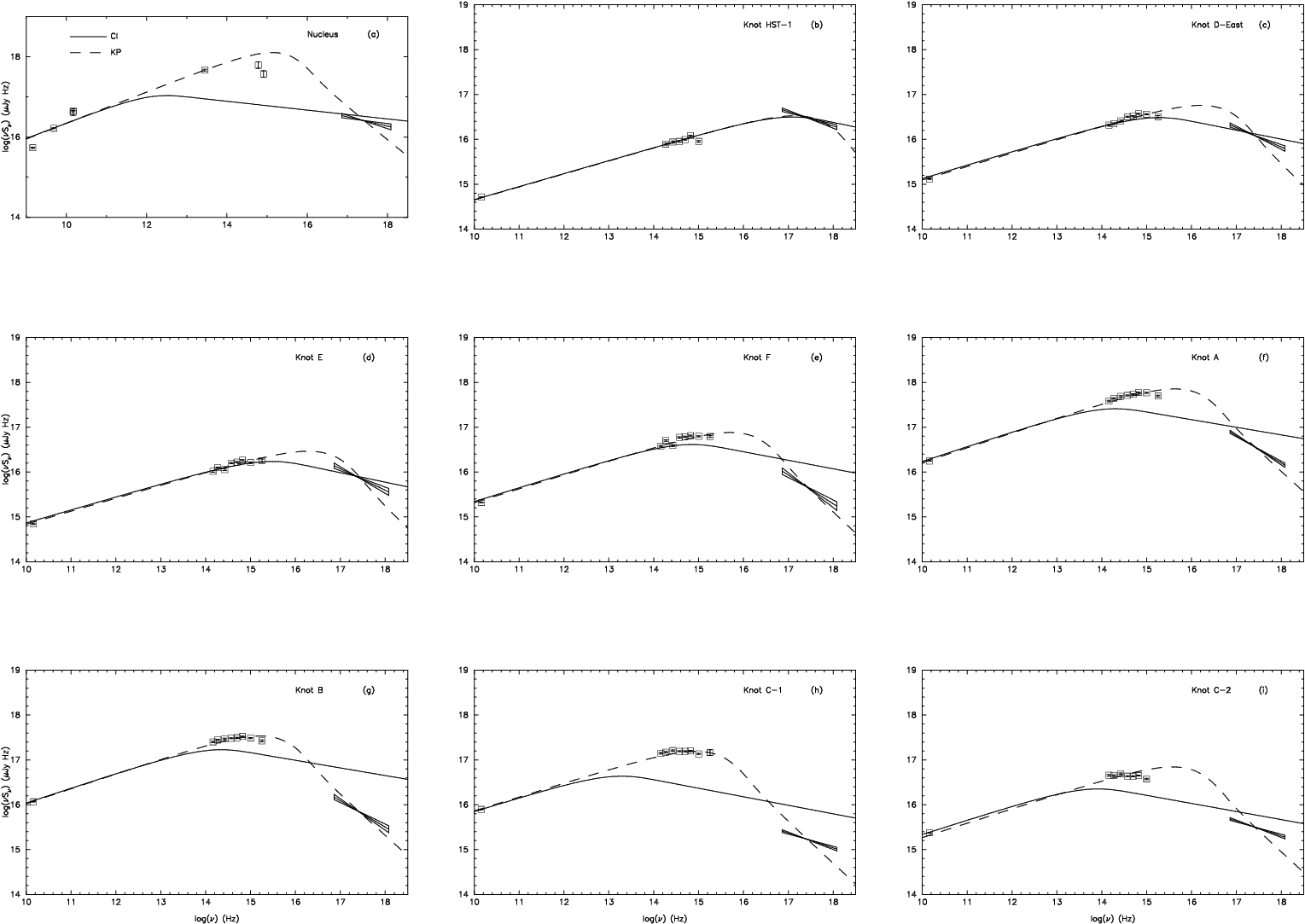}
\caption{CI (solid line) and KP (dashed line) synchrotron spectrum
model fits for nine regions along M87 jet. Data of knots are from:
{\sl VLA}-\citealt{Zhou98}; {\sl HST}-\citealt{Perlman01},
\citealt{Waters05}; {\sl Chandra}-\citealt{Perlman05}. Data of
nucleus are from: {\sl VLA, HST}-\citealt{Sparks96} and references
therein; {\sl Chandra}-\citealt{Perlman05}.}\label{fig1}
\end{figure*}

In Figure~\ref{fig1}, we plot results of CI as well as KP model
fitting for the radio-to-X-ray SED of nucleus and eight knotty
regions. It is worth to notice that knot HST-1 has started to
flare (in radio, optical and X-ray waveband) around 2002
\citep{Harris06}. Since we aim to do 'quiescent state fit' of the
non-simultaneous SED, the broad-band data for this knotty region
of M87 jet are carefully selected to be prior to the flare. The
{\sl VLA} observational data point at 15~GHz is from
\citet{Zhou98}, and optical-near-infrared data points are obtained
with {\sl HST} on 1998 \citep{Perlman01}. As to the X-ray
spectrum, it is from \citet{Perlman05}, which is the reanalysis of
early {\sl Chandra} observations of M87 jet taken on 2000 July
\citep{Wilson02}. The observations above locate prior to the
outburst of HST-1 knot.

Moreover, \citet{Waters05} present new ultraviolet {\sl HST}
observations of M87 jet taken on 2001 February 23. Their results
suggest that knot HST-1 reveals a significant increase in the
brightness from late 2001 to the present. \citet{Waters05} try to
do best fit to the observed radio-to-X-ray SED of HST-1 with
standard synchrotron spectrum models, and found the new UV point
exceeds the fit by a large amount independent of the models. Their
work implies HST-1 knot has started its early evolution of the
flare by the time of their observations. Therefore, we  include
the UV observational data points of \citet{Waters05} for most M87
jet knots (no significant variability found up to date) except
HST-1.

\begin{table*}
\begin{minipage}{126mm}
\caption{Best-Fit Parameters for Synchrotron Spectrum Models}
\label{table1}
\begin{tabular}{@{}lcll}
\hline Jet Region & $\alpha_{\rmn{inj}}$ & CI Model & KP Model \\
\hline Nucleus & 0.70 &
$\nu_{\rmn{B}}=2.22\pm0.12\times10^{12}{\rmn{Hz}}$
& $3.29\pm0.07\times10^{15}{\rmn{Hz}}$ \\
  &  & $\chi^2=119.47$ & 56.05 \\
HST-1 & 0.71 & $\nu_{\rmn{B}}=1.84\pm0.09\times10^{17}$Hz &
$5.91\pm0.19\times10^{17}$Hz \\
  &  & $\chi^2=7.51$ & 5.69 \\
D-East & 0.70 & $\nu_{\rmn{B}}=2.58\pm0.10\times10^{15}$Hz &
$6.44\pm0.08\times10^{16}$Hz \\
  &  & $\chi^2=27.91$ & 36.30 \\
E & 0.71 & $\nu_{\rmn{B}}=4.36\pm0.22\times10^{15}$Hz &
$8.17\pm0.14\times10^{16}$Hz \\
  &  & $\chi^2=15.26$ & 20.83 \\
F & 0.69 & $\nu_{\rmn{B}}=8.95\pm0.43\times10^{14}$Hz &
$1.95\pm0.04\times10^{16}$Hz \\
  &  & $\chi^2=197.93$ & 14.41 \\
A & 0.67 & $\nu_{\rmn{B}}=2.16\pm0.06\times10^{14}$Hz &
$1.42\pm0.02\times10^{16}$Hz \\
  &  & $\chi^2=610.35$ & 18.81 \\
B & 0.67 & $\nu_{\rmn{B}}=2.38\pm0.07\times10^{14}$Hz &
$5.45\pm0.10\times10^{15}$Hz \\
  &  & $\chi^2=764.86$ & 8.58 \\
C-1 & 0.69 & $\nu_{\rmn{B}}=2.42\pm0.06\times10^{13}$Hz &
$3.30\pm0.04\times10^{15}$Hz \\
  &  & $\chi^2=2554.64$ & 59.76 \\
C-2 & 0.68 & $\nu_{\rmn{B}}=9.64\pm0.28\times10^{13}$Hz &
$1.34\pm0.01\times10^{16}$Hz \\
  &  & $\chi^2=508.73$ & 92.74 \\
   \hline
\end{tabular}
\end{minipage}
\end{table*}

Table~\ref{table1} lists the best-fit parameters for synchrotron
spectrum models. During the fitting procedure, we set the input
low frequency spectral index $\alpha_{\rmn{inj}}$ as the observed
radio through optical spectral index $\alpha_{\rmn{ro}}$ (the same
treatment as \citealt{Perlman01}). Here the convention of
synchrotron spectrum is $S_{\nu}\propto\nu^{-\alpha}$.

\begin{figure*}
\includegraphics[angle=-90, totalheight=100mm, keepaspectratio=true]{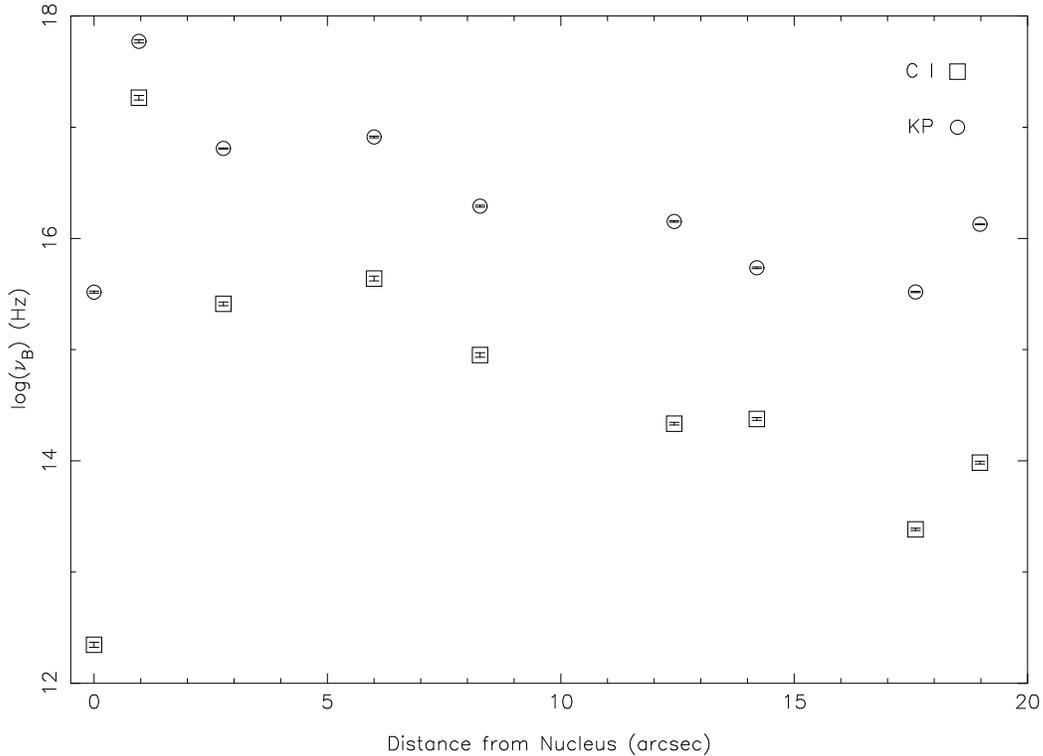}
\caption{Derived break frequency through CI and KP model fitting
versus the projected distance from jet nucleus.} \label{fig2}
\end{figure*}

In Figure~\ref{fig2}, we plot the derived break frequency
$\nu_{\rmn{B}}$ along M87 jet. Our result reveals a general trend
of decline on the break frequency of large-scale knots, similar to
that discovered by \citet{Marshall02}. We find, however, a
significant increase of $\nu_{\rmn{B}}$ from jet nucleus to first
knot, HST-1, by both of the model fitting. With CI model fitting
to mid-IR SED, \citet{Perlman01b} derived break frequency of M87
nucleus, $\nu_{\rmn{B}}=2.8\times10^{12}$Hz. The good agreement
between our derived $\nu_{\rmn{B}}$
($2.22\times10^{12}{\rmn{Hz}}$) of the nucleus based on
multi-waveband fitting  and that of \citet{Perlman01b} suggests
that the increase-decline trend is reliable.

Here we discuss further the origin of broad-band spectra of the
unresolved nucleus of M87 jet. There is some debate on this
problem. \citet{Marshall02} present high resolution X-ray image of
M87 jet using {\sl Chandra}. They find that the core flux is
significantly larger than expected from an advection accretion
flow and the spectrum is much steeper. Their results indicate that
the X-ray emission of the M87 nucleus is due to synchrotron
radiation from a small scale jet. \citet{Wilson02} also give {\sl
Chandra} X-ray imaging and spectroscopy of M87 jet. In view of the
similar spectra of the nucleus and jet knots, and the high X-ray
flux of the knots closest to the nucleus, \citet{Wilson02} suggest
that the X-ray emission of M87 nucleus may actually originate from
the pc- or sub-pc-scale jet rather than the accretion disc. On the
other hand, \citet{DiMatteo03} argue that the observed
radio-to-X-ray spectrum of M87 nucleus is consistent with that
predicted by an radiatively-inefficient accretion disc. However,
they can not also rule out the possibility that the X-ray emission
of the nucleus is dominated by jet emission.

Considering this problem, it is also worthy to notice the latest
research of \citet{Lenain08}. Similar to previous researches,
\citet{Lenain08} suggest that standard one-zone jet model can
describe roughly the radio to X-ray SED of M87 nucleus (see also
figure(1a) of \citet{Lenain08}). They find, however, the single-zone
approach cannot describe correctly the very high energy (VHE)
emission of M87 detected by High Energy Stereoscopic System (HESS)
telescope array. Their further research indicates that a multi-blob
jet model of M87 nucleus may do better at TeV energy band. We note,
however, there is some controversy on the origin of the unresolved
TeV $\gamma$-ray source found in M87. For example, detailed
modelling of \citet{Honda07} suggests that the VHE emissions may
originate from bright knot~A of M87 jet. Therefore, we argue that a
single-zone approach of jet model is tolerable if our research
interests focus only on radio to X-ray SED of M87 nucleus.

If the broad-band emissions of M87 nucleus mainly come from the
unresolved compact jet region, it is necessary also to evaluate
the influence of synchrotron-self-absorption effects at radio
frequencies upon the result of synchrotron model fits. We find out
that the radio spectrum of M87 nucleus is roughly consistent with
a single power-law, i.e. $S_{\nu}\propto\nu^{-0.1}$. The flat
radio spectrum implies that M87 nucleus is only partially opaque
at the observed radio frequencies, and is far beyond the optically
thick regime of synchrotron-self-absorption, i.e.
$S_{\nu}\propto\nu^{2.5}$ \citep{Pacholczyk70}. Consequently, the
synchrotron model fit to radio-to-X-ray SED neglecting the
synchrotron-self-absorption effect should alter the derived break
frequency only slightly, e.g. our derived
$\nu_{\rmn{B,CI}}=2.22\times10^{12}$Hz is in good consistent with
$\nu_{\rmn{B,CI}}=2.8\times10^{12}$Hz derived by
\citet{Perlman01b} based only upon mid-IR SED of M87 nucleus. We
will further indicate that the deduced Dopper beaming factor is
insensitive to the exact value of break frequency (see \S3),
therefore the simplification of broad-band fit to M87 nucleus at
the radio frequencies should not alter the result significantly.

Inspection of Figure~\ref{fig1} and Figure~\ref{fig2} also
indicates that there is systematic difference between CI and KP
model fitting. KP model predicts higher break frequency than CI
model does. Throughout M87 jet, CI model overpredicts the X-ray
flux and the theoretical X-ray spectrum is flatter than that
observed by {\sl Chandra}. {\it Vice versa}, KP model
underpredicts the X-ray flux, and the predicted X-ray spectrum is
steeper than that observed. The same result is also achieved by
\citet{Perlman05}. The systematic deviation of the two models off
the X-ray spectra implies that the real position of
$\nu_{\rmn{B}}$ may deviate also from standard CI and KP cases, or
a modified non-standard synchrotron spectrum model should work
better.

In order to explain such disagreement between the predicted and
observed X-ray spectrum in M87 jet knots, several possible
solutions to this problem are proposed. For example, if
considering synchrotron cooling as well as electron acceleration
processes in a non-uniform magnetic field, \citet{Bicknell96}
suggest it is possible to achieve a larger break in spectral index
than standard CI model, i.e. $\Delta\alpha>0.5$. With a similar
scenario, \citet{Honda07} obtain a modified X-ray spectral index
consistent with {\sl Chandra} observations of M87 knot~A.

During the preparation of this paper, we also notice the latest
result of \citet{Liu07}. They propose a modified CI model to
explain the radio-to-X-ray continua in six knots of M87 jet.
Considering the thin acceleration region (i.e. shock front)
locating at the immediately upstream of the main emission blob,
the broadband spectra can be fit much better than standard CI
model.

\section[]{Deriving Doppler Beaming Factor along M87 Jet}
Through synchrotron spectrum model fitting, we can estimate break
frequency ($\nu_{\rm B}$) of the observed SED. This frequency
provides us with important information of the jet region. In
general, we consider injection of a power-law distribution of
relativistic electrons into an emission region dominated by
synchrotron cooling process. Then the observed break frequency
will correspond to a characteristic break energy of electron
population ($\gamma=\gamma_{\rmn{br}}$), where synchrotron
radiation loss is balanced by escape of electrons from the
emission blob (e.g. \citealt{Inoue96}, \citealt{Kataoka00}).

Therefore we have an relation at the break frequency
\begin{equation}
\label{eq1} t_{\rm syn}(\nu_{\rm B})=t_{\rm esc},
\end{equation}
where $t_{\rm syn}(\nu_{\rm B})$ is synchrotron cooling time at
the break frequency, $t_{\rm esc}$ is diffusive escape time scale
of electrons from the emission region.

The synchrotron cooling time can be further expressed as
\begin{equation}
\label{eq2} t_{\rm syn}(\nu_{\rm
B})=5.08\times10^{16}B^{-3/2}\nu_{\rm B}^{-1/2} {\rm sec},
\end{equation}
where the magnetic field $B$, is in $\mu {\rm G}$, and the break frequency
$\nu_{\rm B}$, is in GHz.

Up to date, we know little about the random motion of electrons
moving in the emission blob. Thus the escape time of electrons,
$t_{\rm esc}$, is quite uncertain. In general, it should be longer
than light travel time over the source ($R/c$, where $R$ is the
size of the emission region and $c$ is speed of light). The escape
time, however, is unlikely to be much longer than $R/c$.
Otherwise, no spectral break ($\nu_{\rm B}$) will be observed in
the SED, which is not the fact of the observations. We then adopt
the treatment of \citet{Kataoka99} as
\begin{equation}
\label{eq3} t_{\rm esc}=\eta R/c, (1<\eta<10)
\end{equation}
where $\eta$ is a dimensionless parameter. To estimate the
influence of escape time upon the uncertainties of Doppler factor,
we further assume that $t_{\rm esc}$ is Gaussian distributed in
this regime (mean value $5.5R/c$, $\pm3\sigma$ range), i.e.
$\eta=5.5\pm1.5$.

The equations above are all written in jet frame. Considering the
relativistic bulk motion of the jet region described with a
certain Doppler beaming factor, $\delta$, there is a Lorentz
transformation of the physical quantities from the source (jet)
frame to the observer frame. According to the formulae presented
in appendixes of \citet{Harris02}, \citet{Stawarz03} and
\citet*{Begelman84}, we substitute (\ref{eq2}) and (\ref{eq3})
into (\ref{eq1}) and rewrite the physical quantities in the
observer frame. Thus Doppler beaming factor appears on both sides
of the relation, which naturally leads to the final formula of
$\delta$ (see Appendix~A for details)
\begin{equation}
\label{eq4} \delta=\left( \frac {\eta R_{\rm obs}}{c}\times
\frac{\nu_{\rm B,obs}^{1/2}B(1)^{3/2}}{1.61\times10^{21}}
\right)^{7/18},
\end{equation}
where $R_{\rm obs}$ is the mean observed radius of emission region
in cm, $c$ is speed of light in ${\rm cm\cdot s^{-1}}$, $\nu_{\rm
B,obs}$ is the observed break frequency in Hz, $B(1)$ is the
equipartition magnetic field in $\mu {\rm G}$ calculated for no
beaming ($\delta=1$). We compute $B(1)$ using eq.~(A6) of
\citet{Harris02}, which requires only total synchrotron luminosity
and $R_{\rm obs}$ of the source. We can fit the radio-to-X-ray
spectrum approximately with a single power law to estimate the
observed bolometric luminosity of the jet region, following the
treatment of \citet{Harris03}. On the other hand, if we have FWHM
measurement of the major axis ($\theta_{\rm a}$) and minor axis
($\theta_{\rm b}$) for a jet region from high resolution imaging
observations, we can use the quantity $\sqrt{\theta_{\rm
a}\times\theta_{\rm b}}$ to estimate the mean angular diameter of
the region, which in turn gives $R_{\rm obs}$ with the known
distance of the object. Due to the extreme proximity of M87 radio
galaxy (distance = 16~Mpc), we neglect the redshift correction in
equation (\ref{eq4}), which requires an additional factor $(1+z)$
multiplied at the right side of that equation.

Although KP model gives better fits to broad-band spectra of M87
jet, this scenario is somewhat unrealistic. The spatial extent of
the X-ray emitting regions in FR~I jets resolved by {\sl Chandra}
implies that continuous acceleration/injection of high-energy
electrons may occur in such objects. Therefore, the break
frequency used in equation ({\ref{eq4}}) to estimate Doppler
factor should be derived from a more realistic modified CI model
(e.g., \citealt{Liu07}). According to the definition of
\citet{Liu07}, their second break frequency, $\nu_{\rmn{B2}}$,
corresponds to the physical break frequency which relates to the
break energy in the spectral energy distribution of relativistic
electrons. They derive $\nu_{\rmn{B2}}$ for five knotty regions
(D, F, A, B, C1), which agree with our $\nu_{\rmn{B,KP}}$ around a
factor of 2. Furthermore, they also report the same decline of
that break frequency down M87 jet.

On the other hand, we find in equation (\ref{eq4}),
$\delta\sim\nu_{\rmn{B}}^{1/5}$, which suggests the value of
Doppler factor is insensitive to exact position of break
frequency. \citet{Liu07} only give physical break frequency for
five knots of M87 jet. To keep a standard treatment, we use
$\nu_{\rmn{B,KP}}$ to approximate the physical break frequency
derived from modified CI model in all nine regions along M87 jet.

\begin{table*}
\begin{minipage}{100mm}
\caption{Related Quantities and Derived Doppler Factors Down M87
Jet} \label{table2}
\begin{tabular}{@{}lllclc}
\hline Jet Region & $\theta_{\rm a}$ (arcsec) & $\theta_{\rm b}$
(arcsec) & $R_{\rmn{obs}}$(pc)& $B(1) (\mu\rmn{G})$ & $\delta$ \\
\hline Nucleus & $0.70\pm0.04$ & $0.58\pm0.03$ & $24.7\pm1.0$ & $699\pm264$ & $2.37\pm0.58$ \\
HST-1 & $0.79\pm0.06$ & $0.60\pm0.04$ & $26.7\pm1.4$ & $237\pm38$ & $3.57\pm0.51$ \\
D-East & $0.85\pm0.06$ & $0.73\pm0.05$ & $30.6\pm1.5$ & $285\pm121$ & $2.72\pm0.74$ \\
E & $1.4\pm0.2$ & $0.9\pm0.1$ & $43.6\pm3.9$ & $183\pm81$ & $2.53\pm0.71$ \\
F & $1.6\pm0.2$ & $1.2\pm0.2$ & $53.8\pm5.6$ & $211\pm143$ & $2.26\pm0.93$ \\
A & $0.98\pm0.04$ & $0.91\pm0.04$ & $36.6\pm1.1$ & $547\pm362$ & $3.18\pm1.27$ \\
B & $1.7\pm0.2$ & $1.3\pm0.2$ & $57.7\pm5.6$ & $319\pm250$ & $2.30\pm1.08$ \\
C-1 & $0.49\pm0.00$ & $0.49\pm0.00$ & $19.0\pm0.0$ & $871\pm668$ & $2.44\pm1.12$ \\
C-2 & $0.49\pm0.00$ & $0.49\pm0.00$ & $19.0\pm0.0$ & $556\pm355$ & $2.46\pm0.95$ \\
 \hline
\end{tabular}

\medskip
$\theta_{\rm a}$ and $\theta_{\rm b}$ are major and minor
component sizes (FWHM), respectively, from {\sl Chandra}
observations \citep{Perlman05}, $R_{\rmn{obs}}$ is mean observed
radius of emission region, $B(1)$ is equipartition magnetic field
calculated for no beaming, i.e. Doppler beaming factor $\delta=1$.
\end{minipage}
\end{table*}

We list in Table~\ref{table2} the related parameters as well as
derived Doppler factors for nine successive regions down M87 jet.
Here $R_{\rmn{obs}}$ is estimated using {\sl Chandra} data of FWHM
component sizes (see Table~1 of \citealt{Perlman05}), for a scale
of 1\arcsec=77.6~pc. We notice that \citet{Kataoka05} calculate
the equipartition magnetic field for three regions of M87 jet,
i.e., knot HST-1, A and D. Their derived values of $B(1)$ which is
based only upon chromatic radio luminosity, agree with our result
within a factor of 2. Therefore our derived parameters are
reliable.

The related parameters are substituted into (\ref{eq4}), then the
values of Doppler beaming factors are derived and listed in last
column of Table~\ref{table2}. The uncertainties of $\delta$ are
calculated using standard theory of error estimation. Our result
provides strong support to the 'modest beaming' scenario (e.g.,
\citealt{Harris03}) in this typical FR~I radio galaxy. It is
worthy to note the new result of \citet{Harris06}. They give the
primary observational result of the long-term {\sl Chandra}
monitoring program of the variable knot HST-1. \citet{Harris06}
report the X-ray decay time of HST-1 for the intensity to drop by
a factor of 2 range from 0.2 to 0.3~yr. Their detailed analysis
suggests that small values of Doppler factor, i.e., $\delta
\approx 3-5$, are favored for this knot. The result of
\citet{Harris06} is well consistent with our independent result
for knot HST-1, i.e., $\delta\approx 3.06-4.08$. Moreover,
\citet{Dodson06} carry out detailed {\sl VSOP} observations toward
the nucleus region of M87 jet. Their study suggests that the M87
nucleus is not strongly Doppler boosted, with a Doppler beaming
factor somewhat larger than 1.6. Our derived $\delta\approx
1.79-2.95$ for nucleus region agree with their work again for this
region. The consistency above then provides strong support to our
new method.

\section[]{Determination of the Intrinsic Velocity Field and Discussions}
Once $\delta$ and the observed apparent speed $\beta_{\rm app}$ of
a jet region are available, we can compute the bulk Lorentz factor
$\Gamma$ and angle to line-of-sight $\theta$ using eq.~(10) and
(11) of \citet{Piner03} as following:
\begin{equation}
\label{eq5} \Gamma=\frac{\beta_{\rm app}^2+\delta^2+1}{2\delta}
\end{equation}
and
\begin{equation}
\label{eq6} \theta=\arctan\frac{2\beta_{\rm app}}{\beta_{\rm
app}^2+\delta^2-1}.
\end{equation}

\begin{table*}
\begin{minipage}{100mm}
\caption{Derived Intrinsic Velocity and Orientation of M87 Jet}
\label{table3}
\begin{tabular}{@{}lcccc}
\hline Jet Region &  $\beta_{\rmn{app}}$ & Reference & $\Gamma$ &
$\theta$ (deg)\\
\hline HST-1 & $6.14\pm0.58$ & B99 & $7.21\pm1.12$ & $13.9\pm1.2$\\
D-East & $2.99\pm0.28$ & B99 & $3.19\pm0.33$ & $21.3\pm5.1$ \\
E & $3.92\pm0.80$ & B99 & $4.50\pm1.36$ & $20.7\pm3.8$ \\
F & $0.86\pm0.23$ & B95 & $1.51\pm0.32$ & $19.5\pm16.0$ \\
A & $0.5090\pm0.0015$ & B95 & $1.79\pm0.56$ & $6.2\pm5.3$ \\
B & $0.62\pm0.05$ & B95 & $1.45\pm0.40$ & $14.9\pm15.1$ \\
C-1 & $0.11\pm0.04$ & B95 & $1.43\pm0.46$ & $2.54\pm2.94$ \\
C-2 & $0.11\pm0.04$ & B95 & $1.44\pm0.40$ & $2.49\pm2.46$ \\
 \hline
\end{tabular}

\medskip
$\beta_{\rmn{app}}$ is observed apparent speed in unit $c$,
$\Gamma$ and $\theta$ refer to bulk Lorentz factor and viewing
angle to line-of-sight of jet flow, respectively. References of
proper motion observations are: B99 \citep*{Biretta99}, B95
\citep*{Biretta95}.
\end{minipage}
\end{table*}

\begin{figure*}
\includegraphics[angle=-90, totalheight=100mm, keepaspectratio=true]{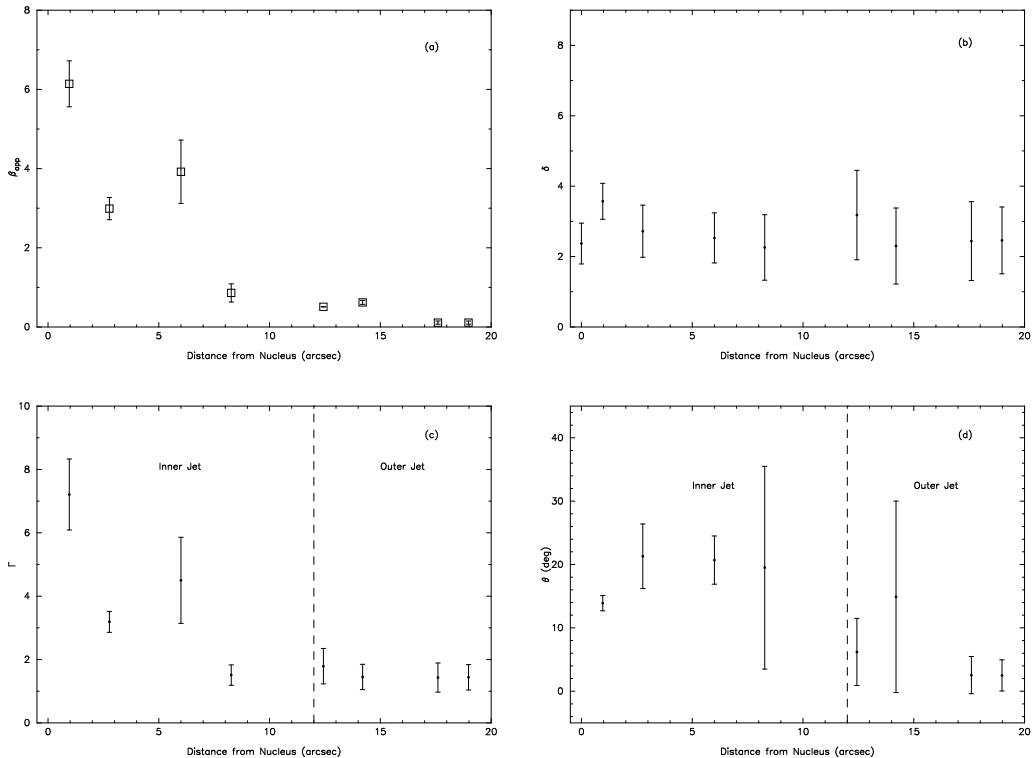}
\caption{Plots of (a) observed apparent speed (in unit $c$), (b)
Doppler beaming factor, (c) intrinsic bulk Lorentz factor and (d)
viewing angle to line-of-sight versus projected distance down M87
jet.} \label{fig3}
\end{figure*}

M87 owns detailed observations of proper motion for most of its
jet knots. In Table~\ref{table3}, we list the observed data of
apparent velocity $\beta_{\rm app}$. Coupling the already derived
Doppler factor $\delta$, the bulk Lorentz factor and viewing angle
of the knotty regions can be calculated with (\ref{eq5}) and
(\ref{eq6}) (see also Table~\ref{table3}). The overall results are
visualized in Figure~\ref{fig3}.

First inspection of Table~\ref{table3} indicates that the viewing
angles of the inner jet regions are fully consistent with
10\degr-19\degr to line-of-sight suggested by previous {\sl HST}
proper motions study of M87 jet \citep{Biretta99}. For HST-1 knot,
we use proper motion data up to $\sim6c$, which are based on
yearly {\sl HST} optical monitoring from 1994 to 1998
\citep{Biretta99}. Recently, \citet{Cheung07} carry out VLBA
monitoring of the same jet region from 2005 to 2006, and report
apparent velocity up to $(4.3\pm0.7)c$, somewhat lower than
previous {\sl HST} measurement of \citet{Biretta99}. Combine this
new apparent velocity with our derived Doppler factor of HST-1
knot, we find $\Gamma=4.51\pm0.85$ and $\theta=15.9\pm1.9$~deg.
Our result agrees well with $\Gamma\geq4.4$ suggested by
\citet{Cheung07}. A comparison between two sets of results of
HST-1 knot in two monitoring intervals suggests that the
orientation of this jet region remains roughly the same, and the
intrinsic velocity of jet flow suffers some decrease. Such a
decrease of bulk velocity of the first knot of M87 jet may imply
some reduced activity of its central engine.

As already noted by many researchers (e.g., \citealt{Perlman01}),
the morphology of the outer M87 jet (knot A, B, C-1, C-2 in
Table~\ref{table3}) is quite different from that of the inner jet
(knot HST-1, D-East, E, F). Inspection of Table~\ref{table3} and
Figure~\ref{fig3} indicates that there is systematic difference on
the intrinsic velocity as well as orientation between these two
segment of jet flow. In Fig.~(\ref{fig3}b), we find the overall
distribution of Doppler factor down M87 jet is quite flat,
supporting the 'modest beaming' scenario. The inner jet, however,
suffers sharp deceleration. On the other side, the intrinsic
velocity of the bulk jet flow remains roughly constant in the
outer jet (see Fig.~(\ref{fig3}c)). In fact, the similar general
trend of $\beta_{\rmn{app}}$ distribution (see Fig.~(\ref{fig3}a))
provides us with some hints.

In general, the early deceleration occurred in typical FR~I jets
such as M87 jet is believed to be caused by continuous mass
loading through entrainment of ambient gaseous medium during jet
propagation (e.g. \citealt{Bicknell94}). Based upon the derived
velocity field here, we will explore the dynamics of deceleration
of M87 jet deeply in a later paper (C. Wang et al., in
preparation) that employs full treatment of special relativity
hydrodynamics.

The orientation of inner jet and outer jet reveals different
pattern also (see last column of Table~\ref{table3} and
Fig.~(\ref{fig3}d). For the inner jet (knot HST-1, D-East, E, F),
our derived viewing angles are fully consistent with
10\degr-19\degr to line-of-sight suggested by the {\sl HST} proper
motion study of \citet{Biretta99} in their Table~3. As to the
outer jet (knot A, B, C-1, C-2), however, we find systematic
deflection to much smaller viewing angles( Despite of the large
uncertainties of viewing angle of knot~B, we infer its most
plausible orientation lies intermediately between that of knot~A
and knot~C-1 considering smooth propagation of jet flow).

Lack of exact measurement of proper motion (e.g., \citet{Ly07}
give VLBI detected apparent speed of core jet as $(0.25-0.40)c$,
while they argue it should be treated only as lower limits), we do
not calculate the intrinsic velocity and orientation for the
nucleus region of M87 jet. We argue, however, the viewing angle of
M87 nucleus should be very close to the first knot, HST-1, i.e.
$\theta\sim15$\degr based upon mean estimation of values of
$\theta\sim14$\degr and 16\degr described above. Using the
observed TeV band spectra of M87, \citet{Lenain08} also constrain
the orientation of its core jet as $\theta\sim15$\degr. Their
results present good support to our viewpoint. Therefore, the idea
that M87 may be a misaligned blazar is favored by our research.

Here we explore further the somewhat strange orientation of the
outer jet. We obtain small viewing angles ($\ll10$\degr) for the
outer parts of M87 jet. The results conflict, however, with much
larger inclination suggested by previous research (e.g.,
\citet{Bicknell96} indicate the orientation of knots A and B
should range between 30\degr-40\degr to line-of-sight). From
equation (\ref{eq4}), we find Doppler factor is a slowly varying
function of the magnetic field, i.e., $\delta\sim B(1)^{7/12}$.
Then the uncomfortable small viewing angles of the outer jet may
be the artifact of the assumed equipartition state of the magnetic
field down the whole outflow. Moreover, we notice the research of
\citet{Heinz97} on M87 jet suggests that there is some departure
from equipartition for the magnetic field of knot~A, i.e.,
$0.2<B/B_{\rm eq}<0.6$. Thus we can further evaluate the effect of
magnetic field values on the derived jet orientation based upon
their constraint to knot~A.

With the constraint to knot~A \citep{Heinz97}, we assume the
magnetic field of this jet region with no beaming, $B(1)$, has
some departure from the related equipartition value (see column~5
of Table~\ref{table2}). Thus we have
\begin{equation}
\label{eq7} B(1)=\xi B_{\rm eq}(1), (0.2<\xi<0.6)
\end{equation}
where $\xi$ is a dimensionless parameter. Using different values
of $\xi$, or different $B(1)$, we then repeat the same procedures
described above to calculate the related Doppler factors, bulk
Lorentz factors and viewing angles. The results are listed in
Table~\ref{table4}.

\begin{table*}
\begin{minipage}{100mm}
\caption{Kinetic Parameters of Knot A with Different Magnetic
Field} \label{table4}
\begin{tabular}{@{}lccc}
\hline $\xi$ &  $\delta$ & $\Gamma$ & $\theta$ (deg)\\
\hline
      0.25 & $1.42\pm0.57$ & $1.15\pm0.11$ & $38.6\pm35.5$ \\
      0.3  & $1.58\pm0.63$ & $1.19\pm0.16$ & $30.1\pm28.2$ \\
      0.4  & $1.87\pm0.75$ & $1.27\pm0.24$ & $20.3\pm19.0$ \\
      0.5  & $2.13\pm0.85$ & $1.36\pm0.31$ & $15.0\pm13.7$ \\
      0.6  & $2.36\pm0.95$ & $1.45\pm0.37$ & $11.9\pm10.7$ \\
 \hline
\end{tabular}
\medskip
\end{minipage}
\end{table*}

Inspection of Table~\ref{table4} suggests that the orientation of
knot~A depends on the level of departure from equipartition. If
$0.4<\xi<0.6$ (small departure), its orientation can maintain the
same viewing angle with the inner jet, i.e.
$\theta\sim10\degr-20\degr$. Meanwhile, larger departure from
equipartition ($0.25<\xi<0.3$) leads to large inclination
($\theta\sim30\degr-40\degr$) consistent with that inferred by
\citet{Bicknell96}. Based on these results, we argue that the
outer jet of M87 starting from knot~A suffers some departure from
equipartition. This departure may result from accumulative mass
entrainment during jet propagation.

\section[]{Conclusions And Summary}
Based on multi-waveband fitting with moderate synchrotron spectrum
models and radiation-escape balance at the break frequency, we
derive the distribution of Doppler beaming factors down M87 jet.
Coupling the data of proper motions, the intrinsic velocity and
orientation are calculated for eight knotty regions, which
presents some hints for the overall dynamics of this famous
extragalactic jet. The main results are summarized as following:

\begin{enumerate}
\renewcommand{\theenumi}{(\arabic{enumi})}
\item The derived distribution of Doppler beaming factors gives
good support to the 'modest beaming' scenario, with $\delta$
varying between $\sim2-5$. \item The bulk flow of M87 jet reveals
sharp deceleration at the inner jet, which may be caused by
continuous entrainment of external gaseous medium during
propagation. Meanwhile, the intrinsic speed of the outer jet
remains roughly constant after early deceleration. \item The
viewing angles of the inner jet regions are fully consistent with
10\degr-19\degr to line-of-sight suggested by previous {\sl HST}
proper motions study of M87 jet. The outer jet, however, shows
systematic deflection off the inner jet to much smaller
inclination ($\theta\ll10$\degr). Further calculation of knot~A
suggests this deflection can be regarded as evidence that the
outer jet suffers some departure from equipartition. \item The
nucleus region of M87 jet should have an viewing angle close to
its first knot HST-1, i.e. $\theta\sim15$\degr. Then our result
favors the idea that M87 may be a misaligned blazar.
\end{enumerate}

We should also comment the recent work of \citet{Laing08}. Using a
relativistic jet model, \citet{Laing08} try to model the high
resolution VLA images of five FR~I radio galaxies other than M87.
They find that a decelerating jet model can fit the observed
brightness and polarization structures well. Moreover, they
suggest the deceleration is most likely caused by mass
entrainment. Their results provide strong support to our result,
i.e. the decelerating velocity profile found in typical FR~I jet
of M87 is reliable.

Once the related jet parameters are derived in
Table~(\ref{table2}) and Table~(\ref{table3}), we can estimate the
contribution to jet power carried by the radiating electrons and
the magnetic field as \citep{Kataoka08}
\begin{equation}
\label{eq8} L_{\rm j}\approx2\pi
R^2c\Gamma^2(\frac{{B'}^2}{8\pi}),
\end{equation}
where the factor 2 means the total internal energy density is
twice of the energy density of equipartition magnetic field.
Making use of derived parameters of HST-1 knot and the
transformation formulae in Appendix~A, we get $L_{\rm
j}\approx9\times10^{43}{\rm erg.s^{-1}}$. Our result agrees well
with $L_{\rm j}\approx10^{44}{\rm erg.s^{-1}}$ suggested by
different methods of previous researches (e.g.,
\citealt{Bicknell96}, \citealt{Stawarz06}).

Another issue we should confirm is whether our derived jet
parameters are consistent with the conservation of the magnetic
flux along M87 jet. In the jet frame, the magnetic flux $S=B'(\pi
{R'}^2)$. Using derived jet parameters in Table~(\ref{table2}) and
transform with formulae of Appendix~A, we can estimate the related
magnetic fluxes (in $\mu {\rm G.pc^2}$) as (the equipartition
level of $B$ in outer jet is somewhat uncertain, thus we only give
results for inner jet regions): $7.23\times10^5$ (Nucleus),
$2.14\times10^5$ (HST-1), $4.10\times10^5$ (D-east),
$5.63\times10^5$ (E), and $1.07\times10^6$ (F). The results
indicate the magnetic fluxes remain roughly conservative down the
jet.

Finally, we should test the validity of main model assumption on
the dominance of synchrotron cooling process. To compare the
dominance of synchrotron cooling rate over SSC (synchrotron
self-Compton) cooling rate, we have the ratio
\begin{equation}
\label{eq9} \frac{{(d\gamma/dt)}_{\rm SSC}}{{(d\gamma/dt)}_{\rm
syn}}=\frac{U'_{\rm ph}}{U'_B},
\end{equation}
where $U'_B={B'}^2/8\pi$ is the intrinsic energy density of
magnetic field. The local energy density of photon field,$U'_{\rm
ph}$, can be estimated with intrinsic total synchrotron luminosity
of the jet region, i.e. $U'_{\rm ph}=L'_s/(4\pi {R'}^2c)$. Making
application to the jet parameters and observed total luminosity of
jet regions, and transform with Appendix~A, we find out the ratio
defined in equation~(\ref{eq9}) range between $10^{-2}-10^{-3}$
along M87 jet. The results are consistent with the model
assumption on the dominance of synchrotron cooling.

Meanwhile, for the inner parts of the jet, the M87 host galaxy may
provide relative intense photon field for the IC scattering process,
resulting in an additional radiative cooling of the
ultra-relativistic electrons. We should evaluate this possible
effect in details. The star light from M87 host galaxy contributes
to the jet as an incident external soft photon field at the optical
band. To simplify the analysis, we assume the external photons have
some characteristic frequency $\nu_{\rm ext}\approx
5\times10^{14}$~Hz. Then after external IC scattering (EC process)
by ultra-relativistic electrons in the jet, the emerged high-energy
photons will have some characteristic frequency as \citep{Fossati98}
\begin{equation}
\label{eq10} \nu_{\rm EC}\simeq
\frac{4}{3}\gamma^2\delta\Gamma\nu_{\rm ext},
\end{equation}
where $\gamma$ is the Lorentz factor of the ultra-relativistic
electron, $\delta$ and $\Gamma$ refer to Doppler beaming factor and
bulk Lorentz factor of the jet flow, respectively.

As to the ultra-relativistic electrons responsible for synchrotron
X-ray emissions in large-scale jets, their typical energy should be
$\gamma\sim 10^6$ or even higher. According to our inferred jet
parameters of M87, we find $\delta\Gamma\sim 10$ for the inner jet
regions. Therefore, the resulting high-energy photons from EC
process locate around $\nu_{\rm EC}\sim 10^{27}$~Hz, which falls in
the TeV energy band. The result above suggests that significant EC
cooling of ultra-relativistic electrons from the host galaxy will
predict significant contribution to the observed TeV spectra of M87.
However, detailed study of \citet{Liu07} implies that the
large-scale knots are unlikely to be the site for the TeV emissions
recently detected in M87. Through modelling of the core jet of M87,
\citet{Lenain08} do not also find evidence of significant
contribution to TeV emissions from EC process of M87 host galaxy.
Therefore, we argue that the synchrotron cooling dominated
assumption of the model should hold in the inner jet regions of M87.

During the preparation of this paper, we notice the latest work of
\citet{Saha08} on M87 jet knots. An improved two-zone (emission
region and nearby acceleration region) CI synchrotron model is
presented by \citet{Saha08}, very similar to the work of
\citet{Liu07}. Both of the modified CI models can fit the
broad-band SED much better than standard synchrotron models for
most of the M87 jet knots. We comment, however, another
possibility for further improvement of the model. All of the
current synchrotron models only consider synchrotron radiation of
relativistic electrons spiralling in an uniform magnetic field,
which is unlikely to be the real case.

It is generally believed that large-scale jet knots are places of
strong shocks, where ultrarelativistic electrons are accelerated via
the first-order Fermi process (e.g., \citealt{Blandford87};
\citealt{Ostrowski02}). There is no strong evidence to support the
suggestion that the magnetic field lines behind the shock front are
straight and uniform. The structure of the magnetic field may be
highly tangled, composed of curved magnetic field lines with
different scales of radius.

On the other hand, theoretical researches on synchrotron radiation
in curved magnetic field lines (\citealt{Zhang95},
\citealt{Cheng96}, \citealt{Zhang98}; the mechanism is named
'synchro-curvature radiation' by the authors) suggest that the
curvature of magnetic field lines plays an important role in the
spectral slope of the synchrotron spectrum, i.e. at the high-energy
portion of the SED. Recently, such a new mechanism was used to
explain successfully the high-energy excess of several gamma-ray
bursts (GRBs; \citealt{Deng05}) the broad-band emission of which may
also originate from relativistic jets. Since standard synchrotron
models show systematic deviation from the observed X-ray spectra of
M87 jet knots, it would be interesting also to explore the effect of
curvature of magnetic field lines on the broad-band SED of the M87
jet and other objects in possible future work.

\section*{Acknowledgments}
We are grateful to Dr.~E.S.~Perlman for fruitful discussions and
for supplying data of M87 jet as well as synchrotron model fitting
code written by Dr.~C.~Carilli. The authors thank a lot the
referee for very careful reading and valuable comments and
suggestions. The authors also thank Dr.~H.L.~Marshall, Y.Y.~Zhou,
J.M.~Bai and D.M.~Meng for helpful discussions. C.C.~Wang
acknowledges Dr.~J.~Kataoka for continuous help on jet physics.
This work is supported by Natural Science Foundation of China
(NSFC) through grant NSFC~10673010 and NSFC~10473013. C.C.~Wang
acknowledges also support from Start-up Fund of USTC. This
research has made use of NASA's Astrophysics Data System.

\appendix

\section[]{Relativistic Transformations to Derive \\* Doppler Beaming Factor}
In this section, we perform relativistic transformations from the
jet rest frame (primed) to the observer frame (unprimed) to derive
Doppler beaming factor of the jet knot. Firstly, let us assume
that the jet knot, observed as a spherical emitting region with
the radius $R$ and the deprojected observed length $L\sim R$, is a
moving blob.

In the jet rest frame, the synchrotron cooling time at the break
frequency can be expressed as \citep{Pacholczyk70}
\begin{equation}
\label{A1} t'_{\rm syn}(\nu'_{\rm B})=5.08\times10^{16}{B'}_{\rm
eq}^{-3/2}{\nu'}_{\rm B}^{-1/2} {\rm sec},
\end{equation}
where the equipartition magnetic field $B'_{\rm eq}$, is in
$\mu$G, and the break frequency $\nu'_{\rm B}$, is in GHz.

While performing Lorentz transformation from jet frame (primed) to
the observer frame (unprimed), the break frequency transforms as
(see formula (C6) of \citealt{Begelman84})
\begin{equation}
\label{A2} \nu'_{\rm B}=\nu_{\rm B,obs}/\delta.
\end{equation}
Meanwhile, we use equation (A8) of \citet{Stawarz03} to transform
the equipartition value of the magnetic field, which corrects the
formula (A7) of \citet{Harris02}. The transformation is expressed
as
\begin{equation}
\label{A3} B'_{\rm eq}=B(1)\delta^{-5/7},
\end{equation}
where $B(1)$ is the equipartition magnetic field calculated for no
beaming ($\delta=1$).

We then substitute equations (\ref{A2}) and (\ref{A3}) into
(\ref{A1}), and the synchrotron cooling time is rewritten as
\begin{equation}
\label{A4} t'_{\rm syn}=5.08\times10^{16}B(1)^{-3/2}\nu_{\rm
B,obs}^{-1/2}\delta^{11/7}.
\end{equation}

On the other hand, in the jet rest frame, $L'=L/\delta \sim
R/\delta$. Therefore we have the volume of the blob $V'=\pi R^2
L'=\pi R^2 L/\delta=V/\delta$. For the assumed geometry of the
knot, the spatial scale of interest in the escape timescale is
$L'$. Then we rewrite the form of escape timescale
\citep{Kataoka99} in the jet rest frame as
\begin{equation}
\label{A5} t'_{\rm esc}=\eta L'/c ,
\end{equation}
where $\eta$ is a dimensionless parameter, and $c$ is speed of
light in ${\rm cm\cdot s^{-1}}$.

The relation above then gives
\begin{equation}
\label{A6} t'_{\rm esc}=\eta \frac{L}{\delta c}\simeq \eta
\frac{R}{\delta c} .
\end{equation}

As to M87 jet, we have FWHM measurement of the major axis
($\theta_{\rm a}$) and minor axis ($\theta_{\rm b}$) for the jet
knots from high resolution imaging observations by {\sl Chandra}
\citep{Perlman05}. Therefore, we can use the quantity
$\sqrt{\theta_{\rm a}\times\theta_{\rm b}}$ (in arcsec) to
estimate the mean angular diameter of the region, which in turn
gives average observed radius $R_{\rm obs}$ with the known
distance of the object. The projection scale of M87 jet is
1\arcsec = 77.6~pc, thus
 \begin{equation}
\label{A7} R_{\rm obs}=0.5\sqrt{\theta_{\rm a}\times\theta_{\rm
b}}\times77.6\times3.09\times10^{18}{\rm cm}.
\end{equation}

At the break frequency of the synchrotron spectrum, the
synchrotron cooling time equals to the escape time of electrons
from the main emission blob (e.g., \citealt{Inoue96},
\citealt{Kataoka00}). When we equal the right side of equation
(\ref{A4}) and (\ref{A6}), and rewrite $\nu_{\rm B,obs}$ in Hz,
the final formula of $\delta$ can be easily expressed as
\begin{equation}
\label{A8} \delta=\left( \frac {\eta R_{\rm obs}}{c}\times
\frac{\nu_{\rm B,obs}^{1/2}B(1)^{3/2}}{1.61\times10^{21}}
\right)^{7/18}.
\end{equation}

 \bsp

\label{lastpage}


\begin{thebibliography}{99}
\bibitem[\protect\citeauthoryear{Begelman, Blandford \& Rees}{1984}]{Begelman84}Begelman M.C., Blandford R.D., Rees M.J., 1984, RvMP, 56, 255
\bibitem[\protect\citeauthoryear{Bicknell}{1994}]{Bicknell94}Bicknell
G.V., 1994, ApJ, 422, 542
\bibitem[\protect\citeauthoryear{Bicknell \&
Begelman}{1996}]{Bicknell96}Bicknell G.V., Begelman M.C., 1996,
ApJ, 467, 597
\bibitem[\protect\citeauthoryear{Biretta, Zhou \&
Owen}{1995}]{Biretta95}Biretta J.A., Zhou F., Owen F.N., 1995,
ApJ, 447, 582
\bibitem[\protect\citeauthoryear{Biretta, Sparks \&
Macchetto}{1999}]{Biretta99}Biretta J.A., Sparks W.B., Macchetto
F., 1999, ApJ, 520, 621
\bibitem[\protect\citeauthoryear{Blandford \& Eichler}{1987}]
{Blandford87}Blandford R., Eichler D., 1987, Phys. Rev., 154, 1
\bibitem[\protect\citeauthoryear{Carilli et al.}{1991}]{Carilli91}Carilli C.L.,
Perley R.A., Dreher J.W., Leahy J.P., 1991, ApJ, 383, 554
\bibitem[\protect\citeauthoryear{Casse \& Marcowith}{2005}]{Casse05}Casse F.,
 Marcowith A., 2005, Astroparticle Physics, 23, 31
\bibitem[\protect\citeauthoryear{Cheng \&
Zhang}{1996}]{Cheng96}Cheng K.S., Zhang J.L., 1996, ApJ, 463, 271
\bibitem[\protect\citeauthoryear{Cheung et
al.}{2007}]{Cheung07}Cheung C.C., Harris D.E., Stawarz \L, 2007,
ApJ, 663, L65
\bibitem[\protect\citeauthoryear{Curtis}{1918}]{Curtis}Curtis H.D., 1918,
Lick Obs. Publ., 13, 31
\bibitem[\protect\citeauthoryear{Deng, Xia \& Liu}{2005}]
{Deng05}Deng X.L., Xia T.S., Liu J., 2005, A\&A, 443, 747
\bibitem[\protect\citeauthoryear{Di Matteo et
al.}{2003}]{DiMatteo03}Di Matteo T., Allen S.W., Fabian A.C.,
Wilson A.S., Young A.J., 2003, ApJ, 582, 133
\bibitem[\protect\citeauthoryear{Dodson, Edwards \&
Hirabayashi}{2006}]{Dodson06}Dodson R., Edwards P.G., Hirabarashi
H., 2006, PASJ, 58, 243
\bibitem[\protect\citeauthoryear{Fossati}{1998}] {Fossati98}Fossati
G., 1998, PhD thesis, SISSA-ISAS
\bibitem[\protect\citeauthoryear{Hardcastle, Birkinshaw \& Worrall}{2001}]{Hardcastle01}Hardcastle M.J., Birkinshaw M.,  Worrall D.M., 2001, MNRAS,
326, 1499
\bibitem[\protect\citeauthoryear{Harris \& Krawczynski}{2002}]{Harris02}Harris
D.E.,  Krawczynski H., 2002, ApJ, 565, 244
\bibitem[\protect\citeauthoryear{Harris et
al.}{2003}]{Harris03}Harris D.E., Biretta J.A., Junor W., Perlman
E.S., Sparks W.B., Wilson A.S., 2003, ApJ, 586, L41
\bibitem[\protect\citeauthoryear{Harris et al.}{2006}]{Harris06}Harris D.E.,
Cheung C.C., Biretta J.A., Sparks W.B., Junor W., Perlman E.S.,  Wilson A.S.,
2006, ApJ, 640, 211
\bibitem[\protect\citeauthoryear{Heavens \& Meisenheimer}{1987}]{Heavens87}Heavens A.,  Meisenheimer K., 1987, MNRAS, 225, 335
\bibitem[\protect\citeauthoryear{Heinz \&
Begelman}{1997}]{Heinz97}Heinz S., Begelman M.C., 1997, ApJ, 490,
653
\bibitem[\protect\citeauthoryear{Honda \&
Honda}{2007}]{Honda07}Honda M., Honda Y.S., 2007, ApJ, 654, 885
\bibitem[\protect\citeauthoryear{Inoue \& Takahara}{1996}]{Inoue96}Inoue S.,
Takahara F., 1996, ApJ, 463, 555
\bibitem[\protect\citeauthoryear{Jaffe \& Perola}{1973}]{Jaffe73}Jaffe W.J.,
 Perola G.C., 1973, A\&A, 26, 421
\bibitem[\protect\citeauthoryear{Kardashev}{1962}]{Kardashev62}Kardashev N.S.,
1962, Soviet Astron.-AJ, 6, 317
\bibitem[\protect\citeauthoryear{Kataoka}{1999}]{Kataoka99}Kataoka J., 1999,
Ph.D. Thesis, University of Tokyo
\bibitem[\protect\citeauthoryear{Kataoka et
al.}{2000}]{Kataoka00}Kataoka J., Takahashi T., Makino F., Inoue
S., Madejski G.M., Tashiro M., Urry C.M., Kubo H., 2000, ApJ, 528,
243
\bibitem[\protect\citeauthoryear{Kataoka \& Stawarz}{2005}]{Kataoka05}Kataoka
J.,  Stawarz \L., 2005, ApJ, 622, 797
\bibitem[\protect\citeauthoryear{Kataoka et al.}{2008}]{Kataoka08}
Kataoka J., Stawarz \L., Harris D.E., Siemiginowska A., Ostrowski
M., Swain M.R., Hardcastle M.J., Goodger J.L., Iwasawa K., Edwards
P.G., 2008, ApJ, 685, 839
\bibitem[\protect\citeauthoryear{Kraft et al.}{2005}]{Kraft05}Kraft R.P.,
Hardcastle M.J., Worrall D.M.,  Murray S.S., 2005, ApJ, 622, 149
\bibitem[\protect\citeauthoryear{Laing \&
Bridle}{2008}]{Laing08}Laing R.A., Bridle A.H, 2008, ASP
Conference Series 386 (Extragalactic Jets: Theory and Observation
from Radio to Gamma Ray), T.A. Rector \& D.S. Yong eds., 70,
arXiv:0801.0147
\bibitem[\protect\citeauthoryear{Leahy}{1991}]{Leahy91}Leahy J.P., 1991, in
Beams and Jets in Astrophysics, ed. P.A.~Hughes (Cambridge: Cambridge Univ.
Press), 100
\bibitem[\protect\citeauthoryear{Lenain et al.}{2008}]{Lenain08}
Lenain J.-P., Boisson C., Sol H., Katarzy\'{n}ski K., 2008, A\&A,
478, 111
\bibitem[\protect\citeauthoryear{Liu \& Shen}{2007}]{Liu07}Liu W.,
Shen Z., 2007, ApJ, 668, L23
\bibitem[\protect\citeauthoryear{Lopez et al.}{2006}]{Lopez06}Lopez L.A.,
Brandt W.N., Vignali C., Schneider D.P., Chartas G.,  Garmire G.P., 2006,
AJ, 131, 1914
\bibitem[\protect\citeauthoryear{Ly, Walker, \&
Junor}{2007}]{Ly07}Ly C., Walker R.C., \& Junor W, 2007, ApJ, 660,
200
\bibitem[\protect\citeauthoryear{Marshall et al.}{2002}]{Marshall02}Marshall
H.L., Miller B.P., Davis D.S., Perlman E.S., Wise M., Canizares C.R.,
Harris D.E., 2002, ApJ, 564, 683
\bibitem[\protect\citeauthoryear{Meisenheimer et al.}{1989}]{Meisenheimer89}Meisenheimer K., Roeser H.-J., Hiltner P., Yates M.G., Longair M.S., Chini R.,
Perley R.A., 1989, A\&A, 219, 63
\bibitem[\protect\citeauthoryear{Myers \& Spangler}{1985}]{Myers85}Myers S.T.,
 Spangler S.R., 1985, ApJ, 291, 52
\bibitem[\protect\citeauthoryear{Ostrowski \& Bednarz}{2002}]
{Ostrowski02}Ostrowski M., Bednarz J, 2002, A\&A, 394, 1141
\bibitem[\protect\citeauthoryear{Pacholczyk}{1970}]{Pacholczyk70}Pacholczyk
A.G., 1970, Radio Astrophysics (San Fransisco: Freeman)
\bibitem[\protect\citeauthoryear{Perlman \& Wilson}{2005}]{Perlman05}Perlman
E.S.,  Wilson A., 2005, ApJ, 627, 140
\bibitem[\protect\citeauthoryear{Perlman et al.}{1999}]{Perlman99}Perlman E.S.,
Biretta J.A., Zhou F., Sparks W.B.,  Macchetto F.D., 1999, AJ, 117, 2185
\bibitem[\protect\citeauthoryear{Perlman et
al.}{2001a}]{Perlman01}Perlman E.S., Biretta J.A., Sparks W.B.,
Macchetto F.D., Leahy J.P., 2001a, ApJ, 551, 206
\bibitem[\protect\citeauthoryear{Perlman et
al.}{2001b}]{Perlman01b}Perlman E.S., Sparks W.B., Radomski J.,
Packham C., Fisher R.S., Pina R., Biretta J.A., 2001b, ApJ, 561,
L51
\bibitem[\protect\citeauthoryear{Piner et al.}{2003}]{Piner03}Piner B.G.,
Unwin S.C., Wehrle A.E., Zook A.C., Urry C.M., Gilmore D.M., 2003, ApJ, 588,
716
\bibitem[\protect\citeauthoryear{Press et al.}{1987}]{Press87}Press W.H.,
Flannery B.P., Teukolsky S.A.,  Vatterling W.T., 1987, in Numerical Recipes
in FORTRAN (Cambridge: Cambridge Univ. Press), 523
\bibitem[\protect\citeauthoryear{Sahayanathan}{2008}]{Saha08}Sahayanathan
S., 2008, MNRAS, 388, L49
\bibitem[\protect\citeauthoryear{Schwartz}{2002}]{Schwartz02}Schwartz
D.A., 2002, ApJ, 569, L23
\bibitem[\protect\citeauthoryear{Sparks, Biretta \&
Macchetto}{1996}]{Sparks96}Sparks W.B., Biretta J.A., Macchetto
F., 1996, ApJ, 473, 254
\bibitem[\protect\citeauthoryear{Stawarz et
al.}{2003}]{Stawarz03}Stawarz \L, Sikora M., Ostrowski M., 2003,
ApJ, 597, 186
\bibitem[\protect\citeauthoryear{Stawarz et
al.}{2006}]{Stawarz06}Stawarz \L, Aharonian F., Kataoka J.,
Ostrowski M., Siemiginowska A., Sikora M., 2006, MNRAS, 370, 981
\bibitem[\protect\citeauthoryear{Waters \&
Zepf}{2005}]{Waters05}Waters C.Z., Zepf S.E., 2005, ApJ, 624, 656
\bibitem[\protect\citeauthoryear{Wilson \& Yang}{2002}]{Wilson02}Wilson A.S.,
 Yang Y., 2002, ApJ, 568, 133
\bibitem[\protect\citeauthoryear{Zhang \&
Cheng}{1995}]{Zhang95}Zhang J.L., Cheng K.S., 1995, Physics
Letters A, 208, 47
\bibitem[\protect\citeauthoryear{Zhang \& Yuan}{1998}]{Zhang98}Zhang J.L.,
Yuan Y.F., 1998, ApJ, 493, 826
\bibitem[\protect\citeauthoryear{Zhou}{1998}]{Zhou98}Zhou F.,
1998, Ph.D. thesis, New Mexico Inst. Mining \& Technology, P99
\end{thebibliography}
\end{document}